\documentclass[prl,showpacs,color,psfig,twocolumn]{revtex4}
\usepackage{amsmath}
\usepackage{amsfonts}
\usepackage[dvips]{graphicx}
\begin{document}

\title{The maximum relative entropy principle}

\author{Jayanth R. Banavar$^{1}$ and Amos Maritan$^{2}$}
\affiliation{$^1$Department of Physics, The Pennsylvania State
University, 104 Davey Laboratory, University Park, Pennsylvania
16802 \\
$^2$Dipartimento di Fisica ``G. Galilei," Universit\`{a} di Padova,
CNISM, via Marzolo 8, 35131 Padova, Italy}

\begin{abstract}
\noindent We show that the naive application of the maximum entropy
principle can yield answers which depend on the level of
description, i.e. the result is not invariant under coarse-graining.
We demonstrate that the correct approach, even for discrete systems,
requires maximization of the relative entropy with a suitable
reference probability, which in some instances can be deduced from
the symmetry properties of the dynamics. We present simple
illustrations of this crucial yet surprising feature in examples of
classical and quantum statistical mechanics, as well as in the field
of ecology.
\end{abstract}

\maketitle

\newcommand{\no}{\noindent}

\newcommand{\mean}[1]{\ensuremath{\left< #1 \right>}}

\newcommand{\de}[1]{\textrm{d}#1}

\newcommand{\diff}[2]{\ensuremath{\frac{\text{d}#1}{\text{d}#2}}}

\newcommand{\ddiff}[2]{\ensuremath{\frac{\text{d}^2#1}{\text{d}#2^2}}}

\newcommand{\pd}[2]{\ensuremath{\frac{\partial #1}{\partial #2}}}

\newcommand{\pdc}[1]{\ensuremath{\frac{\partial}{\partial #1}\,}}

\newcommand{\pdd}[2]{\ensuremath{\frac{\partial^2 #1}{\partial #2 ^2}}}

\newcommand{\pddc}[1]{\ensuremath{\frac{\partial^2}{\partial #1 \,^2}\,}}

There are numerous situations in the natural and social sciences,
medicine, and business which can be described at different levels of
detail in terms of probability distributions. Such descriptions
arise either intrinsically as in quantum mechanics, or because of
the vast amount of details necessary for a complete description as,
for example, in Brownian motion and in many body systems.
Variational methods can be used for constructing an estimate of the
underlying probability distribution when one does not have
sufficient information to deduce it exactly.  The principle of
maximum entropy
\cite{Boltzmann,shannon,jaynes1,jaynes2,Shore,Mead,JaynesBook,dewar1,dewar2,
whitfield} is a widely used variational method for the analysis of
both complex equilibrium and non-equilibrium systems and is being
increasingly employed in a variety of contexts such as image
reconstruction \cite{IMAGE}, crystallography \cite{CRY}, the earth
sciences \cite{ES}, ecology \cite{ECO}, protein science
\cite{werner}, nuclear magnetic resonance spectroscopy \cite{NMR},
X-ray diffraction \cite{xray}, plasma physics \cite{plasma},
condensed matter physics \cite{liquid}, planetary studies
\cite{planet}, electron microscopy \cite{electron}, and neuroscience
\cite{bialek:nature}.

{\bf The maximum entropy principle:} Consider the maximization of
the entropy \cite{shannon}
\begin{equation}\label{ENT}
\mathcal{H}(P) = - \sum_\mathcal{C}P(\mathcal{C}) \ln
P(\mathcal{C}),
\end{equation}
where $ P(\mathcal{C}) $ is the probability that a certain event
$\mathcal{C}$ occurs subject to the constraints:
\begin{equation}\label{C}
\langle Q_r \rangle \equiv \sum_\mathcal{C} P(\mathcal{C})
Q_r(\mathcal{C})\ =\ \bar{ Q_r} \qquad r = 0,1,2, ...    ,
\end{equation}
where the $r$-th constraint requires that the mean value of a
quantity $Q_r$ is equal to $\bar{Q_r}$. $r=0$ is a normalization
condition that ensures that $\sum_\mathcal{C}P(\mathcal{C}) =1 $
and results from selecting $Q_0 = \bar{Q_0} =1$. For $r \geq 1$,
$\bar{ Q_r} $ is obtained from the partial knowledge one has about
the system. The logic underlying the variational method follows
from the link between information and entropy \cite{shannon} - the
more information one has, the lower the entropy. The entropy is
reduced as a result of the partial knowledge encoded in $\bar{
Q_r}$. The entropy maximization principle arises from the
observation that the entropy must be the highest possible that
includes the available information, because a lower entropy would
imply that more information has been incorporated than is
available. Using Lagrange multipliers, $\lambda_r$, to impose the
constraints, one seeks to maximize
\begin{equation}\label{MAX}
- \sum_\mathcal{C}P(\mathcal{C}) \ln P(\mathcal{C}) -\sum_r \
\lambda_r  \ Q_r(\mathcal{C}) \ \ .
\end{equation}
The general solution is found to be
\begin{equation}\label{SOL}
P(\mathcal{C}) = e^{-\sum_r \ \lambda_r  \ Q_r(\mathcal{C})}   ,
\end{equation}
where the $\lambda$'s have to be determined in order to satisfy the
constraints (\ref{C}). In order to illustrate the potential problems
associated with a naive application of the maximum entropy
principle, we begin with a simple classical example. \\

{\bf Application of the maximum entropy principle to classical
statistical mechanics -- derivation of Boltzmann statistics:}
Consider a canonical ensemble of $N$ non-interacting particles that
can occupy discrete, non-degenerate energy levels (the extension to
the degenerate case is straightforward but results in more
cumbersome equations) having an energy $\epsilon_j$, with $j=1,2,3,
....$. We impose an energy constraint: $Q_1 = \sum_j \langle n_j
\rangle \epsilon_j = E$, where $n_j$ is the number of particles in
level $j$ and $\langle\ \cdot\ \rangle$ denotes the average value.
The temperature is a measure of the average total energy. The goal
is to determine the probability, $P$, of observing a given
distribution of particles among the levels subject to a constraint
on the average total energy. For the classical case, the particles
are distinguishable, the identities of the particles are known and
the energy level occupied by the $\alpha$-th particle can be
indicated by $i_{\alpha}$, $\alpha = 1, N$. A straightforward
application of the maximum entropy principle yields the well-known
Boltzmann result: the probability of observing the particle
configuration $\mathbf{i}$
\begin{equation}\label{B1}
P_B(\mathbf{i}) \propto  e^{-\beta \sum_{\alpha=1,N}\sum_j
\delta_{j,i_{\alpha}} \ \epsilon_j}  ,
\end{equation}
where $\mathbf{i}$ denotes the event in which particle $1$ is in
level $i_1$, particle $2$ is in level $i_2$, and so on, the Lagrange
parameter $\beta$ is proportional to the inverse temperature and the
subscript $B$ stands for Boltzmann. In a coarse-grained description
\cite{caticha}, in which one keeps track of just the number of
particles in each level (the occupation number representation), the
relevant event is $\mathbf{n}$, where $n_1$ is the number of
particles in level $1$, $n_2$ is the number of particles in level
$2$, and so on, without regard to their identity. Within the
$\mathbf{n}$ description and starting from Eq. (\ref{B1}), one
obtains for the probability, $P_B^{\prime}(\mathbf{n})$, of the
event $\mathbf{n}$
\begin{equation}\label{B2}
P_B^{\prime}(\mathbf{n}) \propto  \frac {1}{\prod_j n_j!}\ e^{-\beta
\sum_j n_j \ \epsilon_j} ,
\end{equation}
where the prime superscript denotes that the result has been
obtained on coarse-graining.
\\

{\bf Quantum statistics and a puzzle:} The surprising feature of the
maximum entropy principle is that its direct application to the
quantity $P(\mathbf{n})$ yields a result different from Eq.
(\ref{B2}). (The constraints imposed on the system in terms of
$\mathbf{i}$ are also expressible in terms of the coarse-grained
event description $\mathbf{n}$.) One obtains instead the celebrated
Bose-Einstein distribution:
\begin{equation}\label{BE}
P_{BE}(\mathbf{n}) \propto   e^{-\beta \sum_i n_i \ \epsilon_i}
\qquad n_j = 0,1,2,\ldots  .
\end{equation}
On constraining $n_j=0,1$ for all $j$ one gets the Fermi-Dirac
statistics. This result is pleasing because the $\mathbf{n}$
representation is in fact the appropriate one for deriving quantum
statistics. The particles are indistinguishable and all the
information that one has is encapsulated by $P(\mathbf{n})$. The
conundrum is that the results obtained by applying the maximum
entropy principle to $P(\mathbf{i})$ and then coarse-graining the
result to obtain $P(\mathbf{n})$ is different from applying the
maximum entropy principle directly to $P(\mathbf{n})$. In other
words, the operations of maximizing the entropy and of
coarse-graining do {\em not} commute. \\

{\bf  Relative entropy and resolution of the puzzle:}
We suggest that the correct and consistent application of the maximum entropy
principle entails the maximization of the relative entropy
\cite{Kullback} instead of the Shannon entropy in Eq. (\ref{ENT})
subject again to the constraints obtained from partial knowledge
that one has about the system. The relative entropy of $P$
with respect to $P_0$ is defined as \cite{Kullback}
\begin{equation}\label{ENT1}
\mathcal{H}(P|P_0) = - \sum_\mathcal{C}P(\mathcal{C}) \ln \frac
{P(\mathcal{C})} {P_0(\mathcal{C})} \ ,
\end{equation}
where the new term in the denominator $P_0 (\mathcal{C})$ is a
reference term. Such a reference term has been discussed in the
literature in the different context of going from a discrete to a
continuous system and is ``proportional to the limiting density of
discrete points" \cite{JaynesBook}, where it is needed for
dimensional reasons. The reference term is, however, not commonly
invoked as an essential ingredient in the discrete case. It has been
shown by Shore and Johnson \cite{Shore} that ``given a continuous
prior density and new constraints, there is only one posterior
density satisfying these constraints that can be chosen by a
procedure that satisfies the axioms". The unique posterior can be
obtained by maximizing the relative entropy and the axioms pertain
to uniqueness, invariance, system independence and subset
independence. If $ P_0 (\mathcal{C}) $ can be chosen to be a
constant or simply equal to $1$, Eq. (\ref{ENT1}) becomes equivalent
to Eq. (\ref{ENT}). Due to the convexity of the function $x\ln x$,
the relative entropy is never positive and it reaches its maximum
value of zero when $P = P_0$. In the absence of any constraint, the
maximization of the relative entropy leads to the result
$P(\mathcal{C}) = P_0 (\mathcal{C})$.

If the space of events is coarse grained, i.e. it is partitioned
into subsets $\mathcal{C^{\prime}}$, which are pair disjoined,
representing collections of events in $\mathcal{C}$, then the
relative entropy is given by
\begin{equation}\label{ENT2}
\mathcal{H}(P^\prime|P^\prime_0) = -
\sum_{\mathcal{C^\prime}}P^{\prime}(\mathcal{C^\prime}) \ln \frac
{P^{\prime}(\mathcal{C^{\prime}})}
{P_0^{\prime}(\mathcal{C^{\prime}})}  \,
\end{equation}
where the reference term $P_0^{\prime}(\mathcal{C^{\prime}})$ is
obtained straightforwardly by coarse-graining $P_0(\mathcal{C})$ as
\begin{equation}\label{CG}
P_0^{\prime}(\mathcal{C^\prime})\ =\ \sum_{\mathcal{C}\epsilon
\mathcal{C^\prime}}\ P_0(\mathcal{C})  .
\end{equation}
If the constraints Eq. (\ref{C}) are functions only of
$\mathcal{C^\prime}$ then the relative entropy maximization commutes
with the operation of coarse graining and one obtains
\begin{equation}\label{ENTFIN}
P^{\prime}(\mathcal{C^{\prime}})\ =\
P_0^{\prime}(\mathcal{C^\prime}) e^{-\sum_r \ \lambda_r  \
Q_r(\mathcal{C^{\prime}})}     .
\end{equation}

In the derivation of Eq. (\ref{B1}), it was implicitly assumed that
$P_{0,B}(\mathbf{i}) = 1$. On coarse-graining to a description
involving the variable $\mathbf{n}$, Eq. (\ref{CG}) leads to
$P_{0,B}^{\prime}(\mathbf{n}) = N!/\prod_j {n_j!}$ yielding once
again the standard Boltzmann distribution, Eq. (\ref{B2}).  If,
instead, one assumes that $P_{0,BE}(\mathbf{n}) = 1$ then one
derives the Bose Einstein distribution, Eq. (\ref{BE}). \\

{\bf Role of system dynamics:} The success of the principle of
maximum entropy hinges on the choice of the reference probability,
$P_0$, and the identification of the correct constraints not
encapsulated in $P_0$. In the statistical mechanics examples studied
above, the constraint is imposed by fixing the average energy while
the choice of $P_0$ is guided by the postulate that all states are
\textit{a priori} equally probable when one works at the finest
level of description for the system being studied. Of course, this
follows from the dynamics of the system.

Consider the dynamics, in terms of a Markov process, in the
occupation number representation. If the transition rate,
$W^{quantum}(n_j \rightarrow \ n_j+1)$ ($W^{quantum}(n_j \rightarrow
\ n_j-1$)) is proportional to $n_j+1$ ($n_j$) then, in the
stationary state, $P_{0,BE}(\mathbf{n}) = $ constant in agreement
with the implicit choice made for the Bose-Einstein case, Eq.
(\ref{BE}). These transition rates follows from the symmetry of the
quantum wave function describing indistinguishable particles
\cite{BOOK}. For classical (distinguishable) particles, the
transition rate $W^{classical}(n_j \rightarrow \ n_j +1)$ is simply
constant whereas the transition rate $W^{classical}(n_j \rightarrow
\ n_j-1)$ is proportional to $n_j$.  In the stationary state,
$P^{\prime}_{0,B}(\mathbf{n})$ is proportional to $1/\prod_j n_j!$,
which, when used as the reference probability, correctly leads to
Eq. (\ref{B2}). At the description level $\mathbf{i}$, this is
equivalent to $P_{0,B}(\mathbf{i})= $constant. \\

{\bf An ecology application:} A fundamental quantity in ecology is
the probability distribution of the species abundance, i.e. the
probability, $P_{ECO}(\mathbf{n})$, that the first species has a
population $n_1$, the second species, $n_2$ and so on. As an
illustration, consider the simple symmetric case in which all
species are demographically equivalent \cite{Hubbell} and are
governed by similar death and birth rates. The direct application
of the maximum entropy principle without the appropriate
non-trivial reference term and with the constraint that the
average population, $\langle\sum_j n_j\rangle$, is fixed yields a
simple exponential form for the species abundance
\begin{equation}\label{ECO1}
P_{ECO}(\mathbf{n}) \propto e^{-\beta \sum_j n_j}
\end{equation}
The relative species abundance (RSA),$P_{RSA}^{(k)}(n)\equiv
\langle\ \delta_{n,n_k}\rangle$, the probability that the $k$-th
species has population $n$, is thus proportional to $\exp(-\beta
n)$.

In order to choose the reference entropy, we turn to the dynamics as
a guide. Consider a Markov process with transition rates
$W^{eco}(n_j \rightarrow n_j \pm 1)= n_j +c$ where $c$ is a constant
term that, for simplicity, is species independent. When $c = 0$, one
has a simple birth-death process, whose rate is proportional to the
number of individuals of a given species.  A non-zero value of $c$
introduces density dependence in the birth and death rates with a
positive value of $c$ corresponding to a rare-species advantage
\cite{igor2}. The stationary state corresponding to these dynamics
provides a measure of the reference probability
$P_{0,ECO}(\mathbf{n}) \propto \prod_j 1/(n_j+c)$. On applying the
principle of maximum relative entropy with this reference
probability, we find
\begin{equation}\label{ECO2}
P_{ECO}(\mathbf{n}) \propto \prod_j \frac{e^{-\beta n_j}}{n_j+c}
\end{equation}
instead of Eq. (\ref{ECO1}). This leads to a $P_{RSA}^{(k)}(n)
\propto \exp(-\beta n)/(n + c)$. When $c=0$ we obtain the
celebrated Fisher log series \cite{fisher}.  This result can also
be obtained from the standard application of the principle of
maximum entropy by imposing a constraint on the average value of
$\ln n $, a constraint with no ecological basis. When $c$ is
positive,  one obtains the result derived using a density
dependent neutral approach \cite{igor2} which fits the RSA data of
several tropical forests fairly well. The Fisher log-series has a
simple physical interpretation: the $e^{-\beta n}$ term results
from the constraint on the average population whereas the $1/n$
factor follows from the dynamics. The characteristic time scale of
a birth or death event is inversely proportional to $n$, the
number of individuals in a given species -- each individual is a
candidate for dying or for giving birth. The $c$ correction arises
straightforwardly from density
dependence in the birth and/or death rates.\\

{\bf Summary:} The maximum entropy principle is an inference
technique for constructing an estimate of a probability distribution
using available information. We suggest that, in order to guarantee
that the results do not depend on the description level, one ought
to maximize {\em the relative entropy} subject to the known
constraints. This provides a natural interpretation of the relative
entropy \cite{Kullback} in the context of statistical mechanics. In
order to be successful, the method requires knowledge of the
reference probability, which, in turn, depends on the system
dynamics. Alternatively \cite{Mead}, one could maximize the ordinary
entropy $\mathcal{H}(P)$, Eq. (\ref{ENT}), and continue to add
additional constraints until one obtains the correct $P$. In order
to obtain the correct answer, in the absence of the reference
entropy, one requires the knowledge of which optimal constraints to
use (e. g. the constraint on the average value of $\ln n $ in the
ecology illustration) or the use of a large enough number of
constraints \cite{Mead} to ensure convergence. Unfortunately, in
general, there is no \textit {a priori} guarantee that either of
these approaches will be successful.
\\

\textbf{Acknowledgements} We are indebted to Sandro Azaele, Roderick
Dewar, John Harte, Flavio Seno, Antonio Trovato, and Igor Volkov for
insightful discussions. This
work was supported by COFIN 2005 and NSF grant DEB-0346488.\\


\end{document}